\documentstyle[12pt]{article}
\begin{document}
\noindent
\begin{center}
{\Large \bf Refutation of C. W. Misner's claims
in his article \\[1ex]
``Y{\i}lmaz Cancels Newton''}
\\[1em]
Carroll O. Alley and Per Kennett Aschan \\
{\it Department of Physics, University of Maryland \\
College Park, MD 20742-4111, USA} \\
email: {\tt coa@kelvin.umd.edu} \\[2ex]
H\"{u}seyin Y{\i}lmaz \\
{\it Hamamatsu Photonics, K. K., Hamamatsu City 435, Japan \\
Electro-Optics Technology Center, Tufts University \\
Medford, MA 20155} \\[2ex]
30 June 1995
\end{center}

\vspace{1em}
\centerline{\bf Abstract}
\begin{quote}
It is shown that an article by C. W. Misner \cite{r1}
contains serious errors. In particular, the claim that the
Y{\i}lmaz theory of gravitation cancels the Newtonian gravitational
interaction is based on a false premise. With the correct premise
the conclusion of the article regarding the absence of gravitational
interactions applies to general relativity and not to the Y{\i}lmaz theory.
\\[3ex]
PACS 04.20.Cv   -- Fundamental problems and general formalism.\\
PACS 04.50.+h   -- \ldots\ other theories of gravitation.\\
PACS 04.25.$-$g -- Approximation methods; equations of motion.
\end{quote}

\renewcommand{\thefootnote}{\fnsymbol{footnote}}
\footnotetext{UMD PP-95-146; gr-qc/9506082}
\setlength{\arraycolsep}{0.5mm}

\noindent
\\*[3ex]
In an article by C. W. Misner \cite{r1} the expression
referring to the stress-energy of an ideal fluid,
\begin{equation} \label{e10}
\chi_\mu^{\ \nu} =
  (\rho + P) u_\mu u^\nu - P \delta_\mu^{\ \nu} \; ,
\end{equation}
is called
``relativistic matter'', implying that it would be equated
to the matter part $\tau_\mu^{\ \nu}$ in the Y{\i}lmaz field equations
\cite{r2}
\begin{eqnarray}
\frac{1}{2} G_\mu^{\ \nu} & = &
  \tau_\mu^{\ \nu} + t_\mu^{\ \nu} \label{e11} \\
D_\nu G_\mu^{\ \nu} & \equiv & 0 \label{e12} \\
\partial_\nu (\sqrt{-g} \, \tau_\mu^{\ \nu}) & \equiv & 0 \label{e13} \; ,
\end{eqnarray}
which are written in Cartesian coordinates as we shall later
present the solution for $\chi_\mu^{\ \nu}$ in such coordinates.
[See Appendix A for conventions, definitions and coordinate-independent
forms of $\tau_\mu^{\ \nu}$, $t_\mu^{\ \nu}$ and $\partial_\nu$.]

\vspace{1em}
But, of course, calling $\chi_\mu^{\ \nu}$
``relativistic matter''does not make it so, hence one is first obligated
to find out what it is. We can find this out from the first and second
approximation to the Newtonian theory in the equilibrium limit.
\begin{equation}
\chi_\mu^{\ \nu} =
  (\rho + P) u_\mu u^\nu - P \delta_\mu^{\ \nu}
  \: \Rightarrow \: \left(
      \begin{array}{cccc}
        \rho & & & \\
        & -P & & \\
        & & -P & \\
        & & & -P
      \end{array}
    \right) \label{e14} \; .
\end{equation}
We need only three pieces of information for this, namely,
\begin{eqnarray}
g_{00} & \simeq & 1 + 2 \Phi \; , \quad
g_{ij} \simeq -\delta_{ij} (1 - 2 \Phi) \label{e17} \\
\Phi & = & \frac{r^2}{4 R^2} \label{e18} \\
P & = & -\frac{3}{8} \: \frac{r^2 - r_0^{\ 2}}{R^4} \label{e19} \; ,
\end{eqnarray}
where $1/R^2 = 2 \rho / 3$. (Note that the pressure is a second-order
quantity in terms of the energy density $\rho$.) In the Newtonian
approximation these lead to the following relations:
\begin{equation}
\partial_i P \simeq \frac{1}{2} \,
  \partial_i g_{00} \chi^{00} \label{e20}
\end{equation}
for the pressure gradient and
\begin{equation}
\rho \, \partial_i \Phi = -\partial_i P \label{e21} \; ,
\end{equation}
because of the equilibrium. Also,
from (\ref{e17}) we get to first order $\tau_0^{\ 0} \simeq \rho$.

\vspace{1em}
Units are chosen $c = 1$ and $4 \pi G = 1$
to avoid having $c$ and $4 \pi G$ appear in most of the equations.
We use $\chi_\mu^{\ \nu}$ to denote the above expression instead of
$T_\mu^{\ \nu}$ because the use of $T_\mu^{\ \nu}$ can be confusing,
sometimes denoting $T_\mu^{\ \nu} = \tau_\mu^{\ \nu}$ and sometimes
$\frac{1}{2} G_\mu^{\ \nu} = T_\mu^{\ \nu}$. Also we let
(\ref{e14}) represent the equilibrium case, $\rho \, dv^i/dt = 0$,
of a perfect fluid sphere since no solution for the more general
non-equilibrium case is presented by \cite{r1} or by us.

\vspace{1em}
We can now test whether $\chi_\mu^{\ \nu}$ could be $\tau_\mu^{\ \nu}$
as \cite{r1} assumes. In the Newtonian limit we have
$\sqrt{-g} P \simeq P$ to second order, so that
$\partial_\nu (\sqrt{-g} \, \chi_0^{\ \nu}) = 0$ and
\begin{equation} \label{e31}
\partial_\nu (\sqrt{-g} \, \chi_i^{\ \nu})
  = -\partial_i (\sqrt{-g} P)
  \simeq -\partial_i P \not\equiv 0 \; ,
\end{equation}
hence the Freud identity (\ref{e13}) is not satisfied.
This shows that $\chi_\mu^{\ \nu}$ can not possibly be equated to
$\tau_\mu^{\ \nu}$ even in the Newtonian approximation.

\vspace{1em}
Let us next test whether $\chi_\mu^{\ \nu}$ could be
identified with $\tau_\mu^{\ \nu} + t_\mu^{\ \nu}$ in which case it
would satisfy the Bianchi identity (\ref{e11}).
Let us write
\begin{equation}
\chi_\mu^{\ \nu} = \tau_\mu^{\ \nu} + t_\mu^{\ \nu}
\end{equation}
and take the covariant divergence. Using the Freud identity (4)
this gives
\begin{eqnarray}
D_\nu \chi_i^{\ \nu}
  & = & \frac{1}{\sqrt{-g}} \partial_\nu (\sqrt{-g} \, \tau_i^{\ \nu})
    -\frac{1}{2} \, \partial_i g_{\alpha \beta} \, \tau^{\alpha \beta}
    + D_\nu t_i^{\ \nu} \label{e32} \\
  & = & -\frac{1}{2} \, \partial_i g_{\alpha \beta} \, \tau^{\alpha \beta}
    + D_\nu t_i^{\ \nu}
    \simeq -\frac{1}{2} \, \partial_i g_{00} \,
    \tau^{00} + \partial_\nu t_i^{\ \nu} \; .
\end{eqnarray}
In this approximation we may use Eqs. (\ref{e20}) and (\ref{eA3}) of
Appendix A to get
\begin{equation} \label{e33}
D_\nu \chi_i^{\ \nu} =
  -\partial_i P - \rho \, \partial_i \Phi \; ,
\end{equation}
which by (\ref{e21}) verifies the Bianchi identity
\begin{equation}
D_\nu (\frac{1}{2} G_i^{\ \nu}) =
D_\nu \chi_i^{\ \nu} =
  \rho \, \partial_i \Phi - \rho \, \partial_i \Phi \equiv 0 \; .
\end{equation}
Although the identity has been verified to second order for simplicity,
it is valid to all orders.

\vspace{1em}
What we have just proved is that $\chi_\mu^{\ \nu}$, which appears at first
sight to be solely matter stress-energy, is in reality
$\tau_\mu^{\ \nu} + t_\mu^{\ \nu}$, that is, $\chi_\mu^{\ \nu}$ is composed
of both matter and field stress-energies. [This is clearly visible in
the actual solution of the equilibrium problem presented in Appendix B.]
In other words, Misner's premise that $\chi_\mu^{\ \nu}$ is matter alone
and that it would be identified with the matter part $\tau_\mu^{\ \nu}$
of Y{\i}lmaz' theory, is false. Misner's claims in \cite{r1} can not
be valid because the basic premise on which they depend is incorrect.

\vspace{1em}
In principle we can stop here and go no further with \cite{r1}.
However, \cite{r1} makes other incorrect statements, clouding the basic
understanding of the subject. For this reason we add a number of notes.
These notes provide additional information and, where needed, explicit
calculations to indicate that \cite{r1} simply fails to
convey the true situation.

\vspace{5em}
\noindent
{\large \bf Notes}
\\*[3ex]
1. Since Misner considers $\chi_\mu^{\ \nu}$ to be purely matter,
as is clear from the discussion after his Eq. (1.1), he makes the
identification
\begin{equation} \label{e40}
\chi_\mu^{\ \nu} = \tau_\mu^{\ \nu}
\end{equation}
in his Eq. (1.2), which we have earlier shown to be his false premise.
Therefore, according to Misner's Eq. (1.1), Einstein's field equations are
written
\begin{equation}
\frac{1}{2} G_\mu^{\ \nu} = \tau_\mu^{\ \nu} \; .
\end{equation}
Now, to second order in equilibrium we get from the left-hand side of
equation (\ref{e40})
\begin{equation}
\partial_\nu (\sqrt{-g} \chi_i^{\ \nu}) \simeq -\partial_i P \; ,
\end{equation}
whereas from the right-hand side we get by the Freud identity (4)
\begin{equation}
\partial_\nu (\sqrt{-g} \tau_\mu^{\ \nu}) = 0 \; .
\end{equation}
Combining, we find
\begin{equation}
-\partial_i P = 0 \; .
\end{equation}
But with his Eq. (4.4) Misner attributes this result to
Y{\i}lmaz' theory.

\vspace{1em}
On the other hand, as we have earlier shown, starting with the correct
premise $\chi_\mu^{\ \nu} = \tau_\mu^{\ \nu} + t_\mu^{\ \nu}$ one derives
the result
\begin{equation}
-\partial_i P - \rho \, \partial_i \Phi = 0 \; ,
\end{equation}
which Misner attributes to Einstein's theory in Eq. (4.4). Thus
the conclusions that Misner draws from that equation are actually
true in reverse. Ironically, therefore, Misner's conclusions regarding
the absence of gravitational interaction applies to general relativity,
and not to the Y{\i}lmaz theory.

\vspace{1em}
\noindent
2. In Appendix A of \cite{r1} the total stress-energy in the Newtonian
limit is stated as
\begin{equation}
T^{ik}_N = \rho u^i u^k + P^{ik} + t^{ik}_N \; ,
\end{equation}
that is, the Newtonian field stress-energy $t^{ik}_N$ is a
necessary part of the total stress-energy. It is then clear
that, in order for a theory to reduce to the Newtonian theory
in the limit, this field stress-energy must be recovered. This
means that the Newtonian theory must be recovered to first and
second order in the limit (a first-order correspondence is not sufficient)
since $t^{ik}_N$ is a second order quantity. This contradicts
the statement in the abstract of \cite{r1}
that the Newtonian limit would not be affected by the field
stress-energy.

\vspace{1em}
\noindent
3. In the past two of us \cite{r3} have challenged the relativity
community to find an acceptable explanation of the simple
and basic Cavendish experiment using general relativity. So far, we have
not received an adequate response to this challenge. Instead, as in
\cite{r1}, the issue is carried into areas not related
to the force between the two spatially separated bodies
in the Cavendish experiment. As can be seen \cite{r1} does
not provide an explanation of the Cavendish experiment by general
relativity.

\vspace{1em}
\noindent
4. The new theory is in principle a fundamental microscopic theory, and,
as such, deals with particles and waves. In such a theory thermodynamic
properties are to arise from the motions and collisions of the particles
and not from the continuum equations of a classical perfect fluid. The
continuum limit is to be arrived at through statistical averages. It
is, however, gratifying that even without this averaging a continuum
solution to the theory is possible as exhibited in Appendix B.

\vspace{1em}
\noindent
5. It is well known that general relativity has severe problems with
quantization. Recently it has been found that this is due to the
absence of $t_\mu^{\ \nu}$ in the field equations of general relativity.
When $t_\mu^{\ \nu}$ is present as in the Y{\i}lmaz theory, the
gravitational field can be quantized via Feynman's method \cite{r4}.
The reason for this is that in order to quantize a field theory by
Feynman's method one has to have a field lagrangian and one can not
have have field lagrangian without having a field stress energy.
Further, it has been found that the quantized theory is finite.
It appears that with the new theory the dream of generations of
physicists is being realized.

\vspace{1em}
\noindent
6. The article \cite{r1} complains about the time-independent exact
N-particle interactive solution
\begin{equation}
g_{00} = e^{2 \Phi} \; , \quad g_{ij} = -\delta_{ij} \, e^{-2 \Phi}
\end{equation}
\begin{equation}
\Phi = -\sum_A \frac{m_A}{|{\bf x} - {\bf x}_A|}
\end{equation}
of the new theory, saying that if it is exact and independent of time,
then nothing can move. This statement can not be correct because in
the Newtonian theory the Poisson equation has exactly such a solution.
We can therefore do whatever is done in the Newtonian theory to
introduce equations of motion and apply it to the Newtonian limit of the new
theory. If a theory did not have such time-independent N-body
solutions then one would have to worry since the Newtonian theory
has them. The terminology of calling such solutions ``static'' is a
misnomer. They should be called instantaneous-action solutions, corresponding
to the $c \rightarrow \infty$ limit, since the explicit time dependence
of $\Phi$ drops out in this limit.

\vspace{1em}
\noindent
7. A most important feature of the new theory is that the Bianchi
identity is satisfied both by the left-hand side and the
right-hand side of the field equations {\it as an identity}, whereas in general
relativity the left-hand side satisfies the Bianchi identity identically
but the right-hand side does not. We are told that Einstein himself was
aware of this and that is why he many times said ``My equation is like a
house with two wings; the left-hand side is made of fine marble, but
the right-hand side is perishable wood''. It is said that it was his
``dream'' to find a right-hand side that also satisfies the Bianchi
identity, but this was judged to be too difficult or impossible, and
it was given up. Instead, the divergence of the right-hand side is forced
to zero. But then this becomes a {\it constraint} on matter or on the field
(or both), making the theory mathematically overdetermined.

\vspace{1em}
\noindent
8. Finally, let it be understood that we wish to implement Einstein's
concept of gravitation as curved spacetime, in the most satisfactory way.
With the advent of modern symbolic calculation software (Mathematica,
Maple, etc.), available to everyone, we calculate in the
spirit of Leibniz' maxim for the resolution of scientific disputes,
"Calculemus", and go with what we see. All such
calculations (which we invite interested persons to carry out for
themselves) lead us to one single overall conclusion: In order to be
physically and mathematically correct, the conventional dictum,
``the right-hand side of the field equations is all stress-energy,
{\it except} the gravitational field stress-energy'', must be
changed into ``the right-hand side of the field equations is all
stress-energy, {\it including} the gravitational field stress-energy.
Without the inclusion of the gravitational field stress-energy
$t_\mu^{\ \nu}$ there is no interaction between bodies of finite
mass because, as in the Newtonian case, the gravitational force
density is the divergence of $t_\mu^{\ \nu}$.

\vspace{1em}
\noindent
We shall not dwell on other minor misconceptions in \cite{r1}.
Enough has been said of its major misconceptions to indicate that
its conclusions must be taken in the reverse direction.

\vspace{5em}
\setcounter{section}{1}
\renewcommand{\thesection}{\Alph{section}}
\setcounter{equation}{0}
\renewcommand{\theequation}{\Alph{section}.\arabic{equation}}
\noindent
{\large \bf Appendix A. Conventions, terminology, and the Y{\i}lmaz theory}
\\*[3ex]
We define the metric $g_{\mu \nu}$ and the Newtonian
potential $\Phi$ so that to first order $g_{00} = 1 + 2 \Phi$,
$g_{ik} = -\delta_{ik} (1 - 2 \Phi)$ and $\chi_0^{\ 0} = \rho \simeq \sigma$.
This causes the Poisson equation to take the form
\begin{equation} \label{eA1}
\nabla^2 \Phi = \sigma \; ,
\end{equation}
where $\nabla^2$ is the ordinary Laplacian and $\sigma =
\sigma_{\mbox{\it \scriptsize ord}}$ is the ordinary mass density as
in the Newtonian theory. We define the gravitational field
stress-energy in the Newtonian limit as (Misner's $t_\mu^{\ \nu}$
is the negative of ours)
\begin{equation} \label{eA2}
t_\mu^{\ \nu}
  = -\partial_\mu \Phi \, \partial^\nu \Phi
    + \frac{1}{2} \, \delta_\mu^{\ \nu}
    \partial_\lambda \Phi \, \partial^\lambda \Phi \; ,
\end{equation}
from which follows
\begin{equation} \label{eA3}
\partial_\nu t_\mu^{\ \nu}
  = \nabla^2 \Phi \, \partial_\mu \Phi = \sigma \, \partial_\mu \Phi \; .
\end{equation}
These equations refer to the time-independent case, as they will be
applied only to the equilibrium state of a sphere of perfect fluid.
In first order $\sigma \simeq \rho$. The non-equilibrium case can be
treated by letting $v^i \neq 0$, $dv^i/dt \neq 0$ but no solution for
this more general case is presented,
neither by \cite{r1} nor by us, so we must refrain from generalities for
which we have no solutions. For example, in \cite{r1} the expression
$\rho \, dv^i/dt$ frequently appears. To be honest, we should really
set this term to zero. As we will see, there is plenty to discuss
and understand already for the case of equilibrium.

\vspace{1em}
In \cite{r1} the quantity
\begin{equation} \label{eA4}
T^{\mu \nu} = (\rho + P) u^\mu u^\nu - P g^{\mu \nu}
\end{equation}
is introduced, stating that it represents the stress-energy tensor
of matter. We write this expression with mixed indices and denote
it as
\begin{equation} \label{eA5}
\chi_\mu^{\ \nu} = (\rho + P) u_\mu u^\nu - P \delta_\mu^{\ \nu} \; .
\end{equation}
There are two reasons for this. One is that the use of $T^{\mu \nu}$
can be ambiguous, sometimes denoting the right-hand side of
$\frac{1}{2} G_\mu^{\ \nu} = T_\mu^{\ \nu}$ and sometimes denoting the
matter part $\tau_\mu^{\ \nu}$ of the Y{\i}lmaz theory.
The second reason is that we really do not know what $\chi_\mu^{\ \nu}$
is. We would like to call it $\chi_\mu^{\ \nu}$ as if it is an
unknown and find out what it is, as is done in the text.

\vspace{1em}
The equations in the Y{\i}lmaz theory are
\begin{eqnarray}
\frac{1}{2} G_\mu^{\ \nu} & = &
  \tau_\mu^{\ \nu} + t_\mu^{\ \nu} \label{eA6} \\
D_\nu G_\mu^{\ \nu} & \equiv & 0 \label{eA7} \\
\bar{\partial}_\nu (\sqrt{-\kappa} \, \tau_\mu^{\ \nu}) &
\equiv & 0 \label{eA8} \; .
\end{eqnarray}
The definitions of
$\tau_\mu^{\ \nu}$ and $t_\mu^{\ \nu}$ are explicitly given by
\begin{eqnarray}
\tau_\mu^{\ \nu}
  & = & \frac{1}{4 \sqrt{-\kappa}} \, \bar{\partial}_\alpha
      \left[
        {\bf g}^{\alpha \lambda} {\bf g}^{\nu \rho}
          (\bar{\partial}_\rho {\bf g}_{\mu \lambda}
          - \bar{\partial}_\lambda {\bf g}_{\mu \rho})
        + \delta_\mu^{\ \nu}
          \bar{\partial}_\beta {\bf g}^{\beta \alpha}
        - \delta_\mu^{\ \alpha}
          \bar{\partial}_\beta {\bf g}^{\beta \nu}
      \right] \\
t_\mu^{\ \nu} & = & W_\mu^{\ \nu}
  - \frac{1}{2} \delta_\mu^{\ \nu} W_\lambda^{\ \lambda}
\end{eqnarray}
where
\begin{eqnarray}
W_\mu^{\ \nu}
  & = & \frac{1}{8 \sqrt{-\kappa}} {\bf g}^{\nu \lambda}
      \left[
        \bar{\partial}_\lambda {\bf g}_{\alpha \beta}
          \bar{\partial}_\mu {\bf g}^{\alpha \beta}
        - 2 \, \bar{\partial}_\lambda (\sqrt{-\kappa}) \,
          \bar{\partial}_\mu (\frac{1}{\sqrt{-\kappa}})
        - 2 \, \bar{\partial}_\alpha {\bf g}_{\lambda \beta}
          \bar{\partial}_\mu {\bf g}^{\alpha \beta}
      \right] \\
{\bf g}^{\mu \nu} & = & \sqrt{-\kappa} \, g^{\mu \nu} \; , \qquad
{\bf g}_{\mu \nu} = \frac{1}{\sqrt{-\kappa}} \, g_{\mu \nu} \; ,
\end{eqnarray}
where $\bar{\partial}_\nu$ denotes the covariant derivative with
respect to the metric $\eta_\mu^{\ \nu}$ of the local flat space \cite{r5} and
\begin{equation}
\sqrt{-\kappa} = \frac{\sqrt{-g}}{\sqrt{-\eta}} \; .
\end{equation}
If the coordinates are locally Lorentzian then $\bar{\partial}_\nu$ is just
the ordinary derivative $\partial_\nu$ as in the text. To make the exposition
simpler we chose Lorentzian coordinates, although in general $\partial_\nu$
should be replaced everywhere by $\bar{\partial}_\nu$. The virtue of this
procedure is that these definitions and the Freud decomposition
$\frac{1}{2} G_\mu^{\ \nu} = \tau_\mu^{\ \nu} + t_\mu^{\ \nu}$ are
valid using any local flat-space coordinate system.

\vspace{1em}
The concept of the covariant derivative with respect to the coordinates
of a local flat space was introduced by N. Rosen \cite{r6} in 1940.
Rosen later used this concept to formulate his bimetric theory, which is
different from the Y{\i}lmaz theory. Y{\i}lmaz uses it to define the matter
and field stress-energy tensors $\tau_\mu^{\ \nu}$ and $t_\mu^{\ \nu}$ in a
coordinate-independent way. Physically $\tau_\mu^{\ \nu}$ and
$t_\mu^{\ \nu}$ are identified through correspondence to the Newtonian
theory and to special relativity.

\vspace{1em}
It can be seen that the Y{\i}lmaz theory is far easier to work with
than general relativity, especially when the four-index Riemann tensor
$R^\rho_{\mu \nu \sigma}$ is not needed. For then the only things to
calculate are $\tau_\mu^{\ \nu}$ and $t_\mu^{\ \nu}$.
In the usual practice of general relativity one does not pay much
attention to the Freud identity, but it is clear that it plays a
fundamental role in providing an unambigous definition of the matter
and field components of the total stress-energy.

\vspace{5em}
\setcounter{section}{2}
\renewcommand{\thesection}{\Alph{section}}
\setcounter{equation}{0}
\renewcommand{\theequation}{\Alph{section}.\arabic{equation}}
\noindent
{\large \bf Appendix B. Interior solution for the perfect fluid sphere}
\\*[3ex]
{}From what is said above it is clear that $\chi_\mu^{\ \nu} =
(\rho + P) u_\mu u^\nu - P \delta_\mu^{\ \nu}$ is of the form
$\tau_\mu^{\ \nu} + t_\mu^{\ \nu}$, hence it does not belong to
general relativity. If it did, the $t_\mu^{\ \nu}$ would be missing.
For as stated in the introduction of \cite{r1} the formulation of
general relativity is that the right-hand side of the field equations is
``everything {\it except} the gravitational field stress-energy''. The
corresponding statement in the new theory is that the right-hand
side is ``everything {\it including} the gravitational field
stress-energy''. One may ask whether it is true that the usual
Schwarzschild interior solution in reality belongs to the new theory?
The answer is, almost yes but not quite.

\vspace{1em}
We present below an iterative
solution valid to second order for the ideal fluid sphere, which is
compared with the Schwarzschild interior solution evaluated to the
same order. The Schwarzschild metric \cite{r7} and our metric are
both given in Cartesian coordinates to make the comparison
intuitively simple. The solution in the new theory is determined
by the following physical information. We require that
\begin{equation}
\frac{1}{2} G_\mu^{\ \nu} = \chi_\mu^{\ \nu} \; ,
\end{equation}
where $\chi_\mu^{\ \nu}$ has the form (\ref{e14}), and
\begin{equation}
\chi_0^{\ 0} = \rho = \tau_0^{\ 0} + t_0^{\ 0} \; , \quad
\chi_i^{\ j} = -\delta_i^{\ j} P \; ,
\end{equation}
\begin{equation}
\tau_0^{\ 0} = \sigma_{\mbox{\it \scriptsize cov}} =
  \mbox{the covariant Laplacian of } \Phi \; ,
\end{equation}
with the pressure $P$ being given by (\ref{e19}).
%
\begin{eqnarray}
  \mbox{Schwarzschild:} \qquad
  g_{00} & = &
  \displaystyle
  \left[
    \frac{3}{2} \left( \frac{1 - \Phi_0}{1 + \Phi_0} \right)
    - \frac{1}{2} \left( \frac{1 - \Phi}{1 + \Phi} \right)
  \right]^2 \; , \quad
  g_{ij} =
  \displaystyle
  -\delta_{ij} \, \frac{1}{(1 + \Phi)^2} \label{eB1} \\[1em]
  \mbox{Y{\i}lmaz:} \qquad
  g_{00} & = & e^{2(\Phi - 3 \Phi_0)(1 - \Phi)} \; , \quad
  g_{ij} = -\delta_{ij} \, e^{-2 \Phi} \; ,
\end{eqnarray}
where
\begin{equation}
\Phi = \frac{1}{6} \sigma_{\mbox{\it \scriptsize ord}} \, r^2
\end{equation}
\\[1ex]
(Note that $g_{00}$ has the form $e^{2 (\Phi + \Psi)}$, where $\Phi$ is a
function of matter density and $\Psi$ is a function of pressure.)
These metrics give, respectively,
\begin{eqnarray}
    \frac{1}{2} G_\mu^{\ \nu} \hspace{21ex}
    \tau_\mu^{\ \nu} \hspace{16ex} &
    t_\mu^{\ \nu} &
  \\[1em]
    \mbox{Schwarzschild:} \qquad
    \left(
      \begin{array}{c|c}
        \rule[-2mm]{0mm}{0mm}
          \sigma_{\mbox{\it \scriptsize ord}} & 0 \\ \hline
        \rule{0mm}{5mm} 0 & -\delta_i^{\ j} P
      \end{array}
    \right) \:\: = \:\:
    \left(
      \begin{array}{c|c}
         \rule[-2mm]{0mm}{0mm}
          \sigma_{\mbox{\it \scriptsize ord}} + S_0^{\ 0} & 0 \\ \hline
        \rule{0mm}{5mm} 0 & -\delta_i^{\ j} P + S_i^{\ j}
      \end{array}
    \right) \; + \; &
    \left(
      \begin{array}{c|c}
        \rule[-2mm]{0mm}{0mm} t_0^{\ 0} & 0 \\ \hline
        \rule{0mm}{5mm} 0 & t_i^{\ j}
      \end{array}
    \right) &
  \\[2em]
    \mbox{Y{\i}lmaz:} \quad
    \left(
      \begin{array}{c|c}
        \rule[-2mm]{0mm}{0mm}
          \sigma_{\mbox{\it \scriptsize cov}} + t_0^{\ 0} & 0 \\ \hline
        \rule{0mm}{5mm} 0 & -\delta_i^{\ j} P
      \end{array}
    \right) \;\; = \;\;
    \left(
      \begin{array}{c|c}
        \rule[-2mm]{0mm}{0mm}
          \sigma_{\mbox{\it \scriptsize cov}} & 0 \\ \hline
        \rule{0mm}{5mm} 0 & -\delta_i^{\ j} P + S_i^{\ j}
      \end{array}
    \right) \;\;\; + \;\;\; &
    \left(
      \begin{array}{c|c}
        \rule[-2mm]{0mm}{0mm} t_0^{\ 0} & 0 \\ \hline
        \rule{0mm}{5mm} 0 & t_i^{\ j}
      \end{array}
    \right) &
\end{eqnarray}
\begin{eqnarray*}
\rho & \simeq & \sigma_{\mbox{\it \scriptsize cov}} + t_0^{\ 0} \\
\sigma_{\mbox{\it \scriptsize ord}} & \simeq &
  \sqrt{-g} \, \sigma_{\mbox{\it \scriptsize cov}} \\[1ex]
\frac{dP}{dr} & \simeq &
  -\rho \, \frac{d\Phi}{dr} \; ,
\end{eqnarray*}
which are valid to first and the second order. These solutions differ
because in the new theory the $\frac{1}{2} G_\mu^{\ \nu} = \rho$
includes not only the matter density $\sigma_{\mbox{\it \scriptsize cov}}$
but also the energy density of the field, $t_0^{\ 0}$, thus satisfying
the mass-energy correspondence of special relativity.

\vspace{1em}
Note that the material stresses $S_\mu^{\ \nu}$ do not have the same
form as $t_\mu^{\ \nu}$, yet at equilibrium where $\Phi = r^2/4 R^2$
they cancel to form the pressure. This can be seen from
\begin{equation}
\begin{array}{rcl@{\qquad}rcl}
S_0^{\ 0} & = & -(\nabla \Phi)^2 + \Phi \, \nabla^2 \Phi &
  t_0^{\ 0} & = & \displaystyle -\frac{1}{2} (\nabla \Phi)^2 \\[1em]
S_1^{\ 1} & = & \displaystyle -\Phi_{,2}^2 - \Phi_{,3}^2
  + \frac{1}{2} \Phi (\Phi_{,22} + \Phi_{,33}) &
  t_1^{\ 1} & = & \displaystyle -\frac{1}{2} (\Phi_{,2}^2
  + \Phi_{,3}^2 - \Phi_{,1}^2) \\[1em]
S_1^{\ 2} & = & \displaystyle -\Phi_{,1}\Phi_{,2}
  + \frac{1}{2} \Phi \Phi_{,12} &
  t_1^{\ 2} & = & \Phi_{,1} \Phi_{,2}
\end{array} \; .
\end{equation}
All other components are obtainable by symmetry. The essential point is
that $t_\mu^{\ \nu}$ is present and forms a fundamental part
of the total stress-energy as in the new theory.

\vspace{1em}
The second-order solution above can be iterated to higher orders in $\Phi$.
For example, to third order the solution is
\begin{eqnarray}
g_{00} & = & \exp [2(\Phi - 3 \Phi_0)(1 - \Phi) + \frac{4}{3} \Phi^3
  - 6 \Phi^2 \Phi_0 + 4 \Phi \Phi_0^2 ] \\
g_{ik} & = & -\delta_{ik} \, e^{-2 \Phi} \; .
\end{eqnarray}

\vspace{1em}
For the Y{\i}lmaz solution $\tau_\mu^{\ \nu}$ and $t_\mu^{\ \nu}$
have the following expressions:
\begin{equation}
\begin{array}{rcl@{\hspace{2cm}}rcl}
\tau_t^{\ t} & = & \sigma_{\mbox{\it \scriptsize cov}} &
t_t^{\ t} & = & \displaystyle
  -\frac{\sigma_{\mbox{\it \scriptsize cov}}^2}{18} r^2 \\[2ex]
\tau_x^{\ x} & = &
  \displaystyle -\frac{\sigma_{\mbox{\it \scriptsize cov}}^2}{18}
    (3 r_0^2 - 4 r^2 + 2 x^2) &
t_x^{\ x} & = & \displaystyle
  \frac{\sigma_{\mbox{\it \scriptsize cov}}^2}{18}
    (2 x^2 - r^2) \\[2ex]
\tau_x^{\ y} & = & \displaystyle
  -\frac{\sigma_{\mbox{\it \scriptsize cov}}^2}{9} x y &
t_x^{\ y} & = & \displaystyle
  \frac{\sigma_{\mbox{\it \scriptsize cov}}^2}{9} x y \\[2ex]
\tau_x^{\ z} & = & \displaystyle
  -\frac{\sigma_{\mbox{\it \scriptsize cov}}^2}{9} x z &
t_x^{\ z} & = & \displaystyle
  \frac{\sigma_{\mbox{\it \scriptsize cov}}^2}{9} x z \\[2ex]
\tau_y^{\ x} & = & \displaystyle
  -\frac{\sigma_{\mbox{\it \scriptsize cov}}^2}{9} y x &
t_y^{\ x} & = & \displaystyle
  \frac{\sigma_{\mbox{\it \scriptsize cov}}^2}{9} y x \\[2ex]
\tau_y^{\ y} & = & \displaystyle
  -\frac{\sigma_{\mbox{\it \scriptsize cov}}^2}{18}
    (3 r_0^2 - 4 r^2 + 2 y^2) &
t_y^{\ y} & = & \displaystyle
  \frac{\sigma_{\mbox{\it \scriptsize cov}}^2}{18}
    (2 y^2 - r^2) \\[2ex]
\tau_y^{\ z} & = & \displaystyle
  -\frac{\sigma_{\mbox{\it \scriptsize cov}}^2}{9} y z &
t_y^{\ z} & = & \displaystyle
  \frac{\sigma_{\mbox{\it \scriptsize cov}}^2}{9} y z \\[2ex]
\tau_z^{\ x} & = & \displaystyle
  -\frac{\sigma_{\mbox{\it \scriptsize cov}}^2}{9} z x &
t_z^{\ x} & = & \displaystyle
  \frac{\sigma_{\mbox{\it \scriptsize cov}}^2}{9} z x \\[2ex]
\tau_z^{\ y} & = & \displaystyle
  -\frac{\sigma_{\mbox{\it \scriptsize cov}}^2}{9} z y &
t_z^{\ y} & = & \displaystyle
  \frac{\sigma_{\mbox{\it \scriptsize cov}}^2}{9} z y \\[2ex]
\tau_z^{\ z} & = & \displaystyle
  -\frac{\sigma_{\mbox{\it \scriptsize cov}}^2}{18}
    (3 r_0^2 - 4 r^2 + 2 z^2) &
t_z^{\ z} & = & \displaystyle
  \frac{\sigma_{\mbox{\it \scriptsize cov}}^2}{18}
    (2 z^2 - r^2) \; ,
\end{array}
\end{equation}
where
\begin{eqnarray}
\sigma_{\mbox{\it \scriptsize ord}} & = & \nabla^2 \Phi \\
\sigma_{\mbox{\it \scriptsize cov}} & = &
  - \frac{1}{\sqrt{-g}} \, \partial_\nu (\sqrt{-g} \, \partial^\nu \Phi) \\
  & \simeq & \frac{1}{\sqrt{-g}} \nabla^2 \Phi \\
  & \simeq & \sigma_{\mbox{\it \scriptsize ord}} \, (1 + 2 \Phi) \; .
\end{eqnarray}

\vspace{1em}
For completeness the corresponding exterior solutions are shown below:
%
\begin{eqnarray}
  \mbox{Schwarzschild:} \qquad
  g_{00} & = &
    \displaystyle
    \left(
      \frac{\displaystyle 1 + \frac{\Phi}{2}}
      {\displaystyle 1 - \frac{\Phi}{2}}
    \right)^2 \; , \quad
  g_{ij} =
    \displaystyle
    -\delta_{ij} \left(
      1 - \frac{\Phi}{2}
    \right)^4 \\[1em]
  \mbox{Y{\i}lmaz:} \qquad
  g_{00} & = & e^{2 \Phi} \; , \quad
  g_{ij} = -\delta_{ij} \, e^{-2 \Phi} \; ,
\end{eqnarray}
where
\begin{equation}
\Phi = -\frac{M}{r}
\end{equation}
\\[1ex]
These metrics give, respectively,
\begin{eqnarray}
    \frac{1}{2} G_\mu^{\ \nu} \qquad \;\; &
    \tau_\mu^{\ \nu} & \qquad \quad
    t_\mu^{\ \nu}
  \\[1em]
    \mbox{Schwarzschild:} \quad \;\;
    \left(
      \begin{array}{c|c}
        \rule[-2mm]{0mm}{0mm} 0 & 0 \\ \hline
        \rule{0mm}{5mm} 0 & 0
      \end{array}
    \right) \;\; = \;\; &
    \left(
      \begin{array}{c|c}
        \rule[-2mm]{0mm}{0mm} 0 & 0 \\ \hline
        \rule{0mm}{5mm} 0 & 0
      \end{array}
    \right) & \;\; + \;\;
    \left(
      \begin{array}{c|c}
        \rule[-2mm]{0mm}{0mm} 0 & 0 \\ \hline
        \rule{0mm}{5mm} 0 & 0
      \end{array}
    \right)
  \\[1em]
    \mbox{Y{\i}lmaz:} \quad
    \left(
      \begin{array}{c|c}
        \rule[-2mm]{0mm}{0mm} t_0^{\ 0} & 0\\ \hline
        \rule{0mm}{5mm} 0 & t_i^{\ j}
      \end{array}
    \right) \; = \; &
    \left(
      \begin{array}{c|c}
        \rule[-2mm]{0mm}{0mm} 0 & 0 \\ \hline
        \rule{0mm}{5mm} 0 & 0
      \end{array}
    \right) & \; + \;
    \left(
      \begin{array}{c|c}
        \rule[-2mm]{0mm}{0mm} t_0^{\ 0} & 0 \\ \hline
        \rule{0mm}{5mm} 0 & t_i^{\ j}
      \end{array}
    \right)
\end{eqnarray}
where
\\[1ex]
\begin{equation}
\begin{array}{rcl}
  t_0^{\ 0} & = & \displaystyle -\frac{1}{2} (\nabla \Phi)^2 \\[1em]
  t_1^{\ 1} & = & \displaystyle -\frac{1}{2} (\Phi_{,2}^2
    + \Phi_{,3}^2 - \Phi_{,1}^2) \\[1em]
  t_1^{\ 2} & = & \Phi_{,1} \Phi_{,2}
\end{array} \; .
\end{equation}
\\[1ex]
All other components are obtainable by symmetry. Note that in the
Schwarzschild solution $t_\mu^{\ \nu}$ is missing.

\vspace{1em}
For the Y{\i}lmaz solution, $\tau_\mu^{\ \nu}$ and
$t_\mu^{\ \nu}$ have the following expressions:
\begin{equation}
\begin{array}{rcl@{\hspace{2cm}}rcl}
\tau_t^{\ t} & = & 0 &
  t_t^{\ t} & = & \displaystyle -\frac{M^2}{2 r^4} \\[2ex]
\tau_x^{\ x} & = &
  \displaystyle 0 &
 t_x^{\ x} & = & \displaystyle \frac{M^2}{2 r^6} (2 x^2 - r^2) \\[2ex]
\tau_x^{\ y} & = & \displaystyle 0 &
  t_x^{\ y} & = & \displaystyle \frac{M^2}{r^6} x y \\[2ex]
\tau_x^{\ z} & = & \displaystyle 0 &
  t_x^{\ z} & = & \displaystyle \frac{M^2}{r^6} x z \\[2ex]
\tau_y^{\ x} & = & \displaystyle 0 &
  t_y^{\ x} & = & \displaystyle \frac{M^2}{r^6} y x \\[2ex]
\tau_y^{\ y} & = & 0 &
  t_y^{\ y} & = & \displaystyle \frac{M^2}{2 r^6} (2 y^2 - r^2) \\[2ex]
\tau_y^{\ z} & = & \displaystyle 0 &
  t_y^{\ z} & = & \displaystyle \frac{M^2}{r^6} y z \\[2ex]
\tau_z^{\ x} & = & \displaystyle 0 &
  t_z^{\ x} & = & \displaystyle \frac{M^2}{r^6} z x \\[2ex]
\tau_z^{\ y} & = & \displaystyle 0 &
  t_z^{\ y} & = & \displaystyle \frac{M^2}{r^6} z y \\[2ex]
\tau_z^{\ z} & = &
  \displaystyle 0 &
  t_z^{\ z} & = & \displaystyle \frac{M^2}{2 r^6} (2 z^2 - r^2) \; ,
\end{array}
\end{equation}

Incidentally, in the Y{\i}lmaz theory there are
no black holes in the sense of event horizons, but there can
be stellar collapse as observed. Radially directed light can always
escape, although red-shifted. However, there are no point
singularities since the invariant curvature quantities for the exterior
such as the Ricci invariant $R = R_\mu^{\ \mu} = 2M^2/(r^4 e^{2M/r})$
and the Kretschmann invariant
$R_{\mu \nu \rho \sigma} R^{\mu \nu \rho \sigma} =
4M^2 (7M^2 - 16Mr +12r^2)/(r^8 e^{4Mr})$ do not diverge. In fact,
they go to zero as the radius goes to zero. This seems to mean that
even in stellar collapse and in the early universe a quantum theory
of gravitation based on Y{\i}lmaz theory will not lead to inconsistency.

\vspace{1em}
Note that if we hold the statement of general relativity as ``the
right-hand side is everything {\it except} field stress-energy'',
neither of the interior solutions discussed above belongs to
general relativity because, contrary to the statement above, the
field stress-energy $t_\mu^{\ \nu}$ is fully present in both.
The main difference between these solutions is that in the
Schwarzschild solution $\frac{1}{2} G_0^{\ 0} =
\sigma_{\mbox{\it \scriptsize ord}}$, whereas in the Y{\i}lmaz theory
$\frac{1}{2} G_0^{\ 0} = \sigma_{\mbox{\it \scriptsize cov}} + t_0^{\ 0}$.
This shows that only the latter solution has the correct
special-relativistic correspondence.

\vspace{1em}
In the Y{\i}lmaz theory the exterior solution
has also a field-energy density $t_0^{\ 0}$. Since
$\Phi_{\mbox{\it \scriptsize int}} = r^2/4 R^2$,
$\Phi_{\mbox{\it \scriptsize ext}} = -M/r$ and
$t_0^{\ 0} = -(\nabla \Phi)^2/(8 \pi)$, we find that
\begin{eqnarray}
{\cal E}
  & = & \int_0^{r_0} t_0^{\ 0} (\mbox{int.}) \, dV
    + \int_{r_0}^\infty t_0^{\ 0} (\mbox{ext.}) \, dV \\[1ex]
  & = & -\frac{M^2}{10 r_0} - \frac{M^2}{2 r_0}
    = -\frac{3}{5} \frac{M^2}{r_0} \; ,
\end{eqnarray}
which is exactly the Newtonian field energy of the fluid sphere.
The interior Schwarzschild metric does not belong to general relativity
since the right-hand side of the field equations includes field
stress-energy, contrary to the statement of general relativity.

\vspace{1em}\
Thus in the Y{\i}lmaz theory the interior and the exterior field
energies are as in the Newtonian theory, whereas in general relativity
there is no field energy in the exterior. Clearly general
relativity does not have a unique correspondence with the Newtonian
theory. For in the exterior case $t_\mu^{\ \nu}$ is assumed to be zero
(in fact, set to zero by $\frac{1}{2} G_\mu^{\ \nu} = 0)$, whereas in
the interior no such condition exists, hence $t_\mu^{\ \nu}$
inadvertently sneaks in. Thus there is an inconsistency between the
exterior and interior solutions in general relativity, whereas in
the Y{\i}lmaz theory the expressions of $\tau_\mu^{\ \nu}$ and
$t_\mu^{\ \nu}$ are unambiguously defined both in the exterior and
the interior (see Appendix A).

\end{document}